\documentclass{article}
\usepackage{spconf,amsmath,graphicx,hyperref}
\usepackage{amssymb}
\usepackage{float}
\usepackage{booktabs}
\usepackage{xcolor}

\title{%SPARO: Sparse 
HRTF Upsampling and Personalization via Neural Operators % Learning
}
\name{Meenakshi Krishnan$^{1}$\sthanks{Work done while at Dolby Laboratories. \mbox{mkrishn9@umd.edu}}, You Zhang$^{2}$, Ramani Duraiswami$^{1}$, C. Phillip Brown$^{2}$}
\address{$^{1}$University of Maryland, College Park $\qquad$ $^{2}$Dolby Laboratories, Inc.}

\begin{document}
\ninept
\maketitle
\begin{abstract}
Head-related transfer function (HRTF) spatial upsampling from sparse measurements is a fundamental challenge in personalized binaural audio. We formulate HRTF upsampling as an operator learning problem and introduce SPARO, a neural operator that maps sparse measurements at a set of spatial directions to the HRTF field, a continuous representation that can be queried at arbitrary positions. SPARO projects the measurements onto a learned set of spherical latent modes, processes the latent modes with Transformer blocks, and decodes the resulting representations at query directions. On SONICOM, using the LAP24 measurement configuration, SPARO attains the best log-spectral distortion (LSD) and interaural level difference (ILD) error at every sparsity level from 3 to 100 measurements (18\% lower LSD and 43\% lower ILD error than the strongest baseline at 100 measurements) while remaining competitive in interaural time difference (ITD) error. It reconstructs a dense HRTF $15$--$373\times$ faster than the leading neural-field method RANF. We further validate the approach on HUTUBS, demonstrating its effectiveness across HRTF datasets.

\end{abstract}
\section{Introduction}
\label{sec:intro}

Personalized spatial audio rendering for virtual and augmented reality (VR/AR) requires head-related transfer functions (HRTFs), which capture the direction-dependent acoustic filtering of a listener's head, torso, and pinnae. As these filtering effects are highly anatomy-specific, the use of non-individualized HRTFs measurably degrades sound localization accuracy and perceptual attributes such as externalization and immersion \cite{wenzel1993localization, moller1996binaural}. Personalized HRTFs are also relevant for hearing assistive devices (including hearing aids and cochlear implant processors) where accurate spatial rendering can improve speech intelligibility in multi-talker environments \cite{cuevas2021impact}.

\iffalse
While acoustic measurement of a full individual HRTF would be the gold standard, this requires impulse responses at hundreds of spatial positions; the SONICOM dataset \cite{engel2023sonicom}, for instance, covers 793 directions per subject, making direct measurement costly and impractical at scale.
\fi

While acoustic measurement remains the gold standard for individual HRTFs, densely sampled datasets typically require impulse responses at hundreds of spatial directions, for example, SONICOM~\cite{engel2023sonicom}, CIPIC \cite{algazi2001cipic}, and ARI \cite{majdak20103} contain 793, 1250, and 1550 measured directions, respectively. Such acquisition is costly and impractical at scale, while practical rendering may additionally require HRTFs at arbitrary directions outside the measurement grid.
HRTF spatial upsampling addresses this bottleneck by estimating dense HRTFs from a sparse set of individual measurements \cite{hogg2025listener, zhong2014and}. 

\begin{figure}[h]
    \centering
    \includegraphics[width=\linewidth, trim={0cm 0cm 0 0},clip]{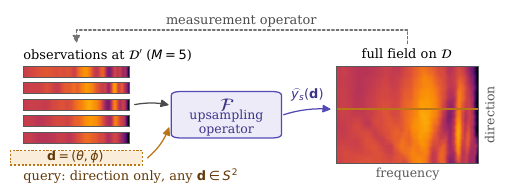}
    \vspace{-7mm}
    \caption{HRTF upsampling as operator learning. SPARO learns
the upsampling operator $\mathcal{F}$ that generalizes across listeners and can be queried at arbitrary directions to produce the HRTF field.} \vspace{-6mm}
    \label{fig:grids0}
\end{figure} 
\begin{figure*}[h]
    \centering
\includegraphics[trim={0cm 3mm 0 0mm},width=0.85\textwidth,clip]{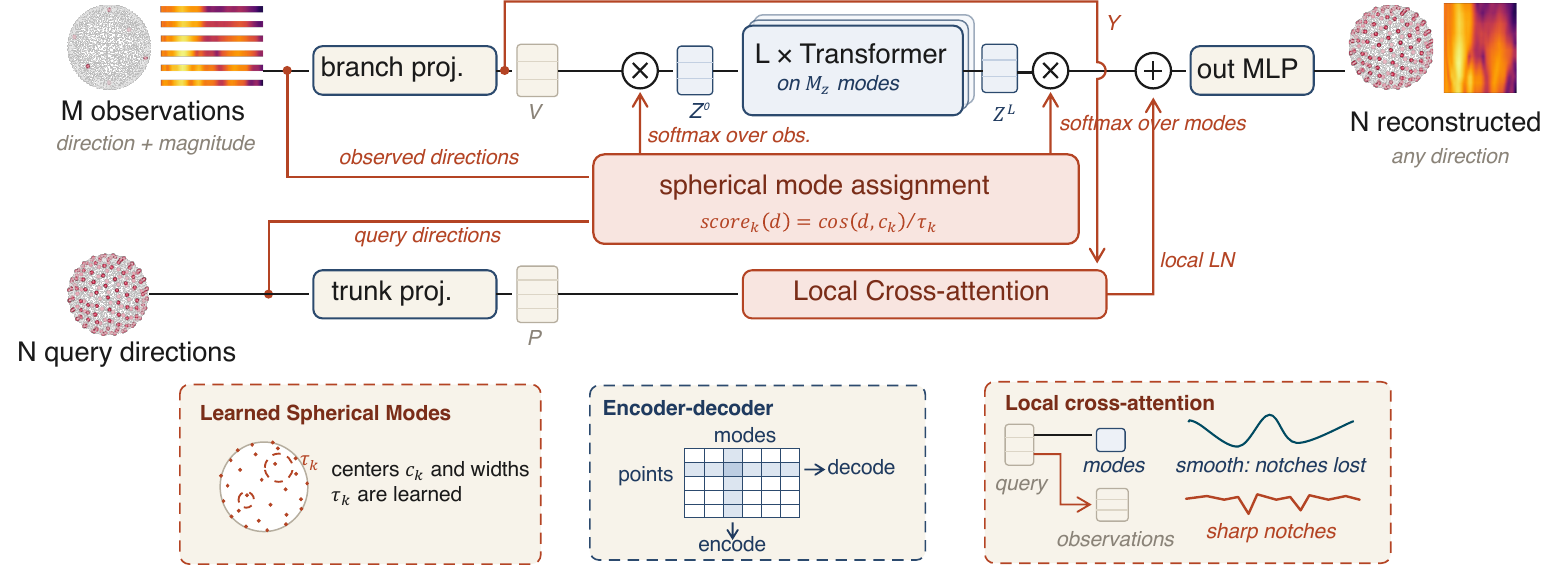}
\vspace{-4mm}
    \caption{Architecture for HRTF upsampling. Sparse HRTF observations are embedded by the branch projector and aggregated into a fixed set of learned spherical modes according to their angular similarity to mode centers on \(S^2\). Transformer blocks model interactions entirely in the latent mode space. At arbitrary query directions, the transformed modes are interpolated using the same spherical mode geometry, and a final MLP produces the output HRTF.} \vspace{-4mm}
    \label{fig:placeholder}
\end{figure*}

Algorithmic upsampling methods such as barycentric interpolation and spherical harmonic decomposition \cite{duraiswami2004interpolation,arend2023magnitude, zotkin2009regularized} are effective when measurements are sufficiently dense, but degrade severely under high sparsity \cite{masuyama2024niirf, masuyama2025retrieval}. Learning-based approaches have shown strong performance for HRTF upsampling from sparse measurements~\cite{hogg2025listener, fantini2025survey}. In particular, neural field (NF) methods represent HRTFs as a function of sound source direction conditioned on subject-specific parameters, enabling a pre-trained generic model to be adapted to new subjects from sparse measurements~\cite{masuyama2024niirf, zhang2023hrtf, masuyama2025sudafield, hu2026graph}. The Retrieval-Augmented Neural Field (RANF)~\cite{masuyama2025retrieval}, the winning system of Task 2 of the Listener Acoustic Personalization (LAP24) challenge~\cite{hogg2025listener}, further augments this with similar-subject retrieval, achieving state-of-the-art performance in log-spectral distortion (LSD), interaural level difference (ILD) and interaural time difference (ITD) errors across sparsity levels of 3 to 100 measurements.

%Many of these NF-based approaches use per-subject iterative adaptation at inference — typically optimizing low-rank subject-specific parameters \cite{masuyama2024niirf, masuyama2025retrieval,hu2022lora} on the available sparse measurements. 

Neural-field approaches provide a flexible continuous representation of HRTFs~\cite{zhang2023hrtf}, where personalization typically involves optimizing subject-specific representations or parameters from the available measurements. Many methods thus use per-subject iterative adaptation at inference, for example by optimizing low-rank subject-specific parameters \cite{masuyama2024niirf,masuyama2025retrieval}. In this formulation, the sparse measurements primarily serve to identify a listener-specific field. Rather than representing each listener's HRTF as a separate field, an alternative is to learn a single upsampling operator that maps sparse HRTF measurements to the corresponding HRTF field across listeners.

In general, operator learning methods aim to learn mappings between infinite-dimensional function spaces, with the functions represented in practice by their values on finite sets of sample locations \cite{kovachki2023neural,lu2021learning}. They offer a discretization-invariant representation that is independent of the input-output dimensions, and prior architectures \cite{wang2024latent, krishnan2026gaia} decouple input and output discretizations through a fixed latent representation. In this paper, we learn a shared upsampling operator across individuals that maps sparse measurements to the HRTF at arbitrary query directions, as illustrated in Fig.~\ref{fig:grids0}.

% Operator learning methods learn mappings between infinite dimensional function spaces from data, and are used to solve problems governed by Partial Differential Equations (PDEs). Rather than solving per instance \cite{raissi2019physics}, a trained operator learning model generalizes across a family of problem instances by approximating the underlying solution operator directly from observations~\cite{kovachki2023neural, lu2021learning}. HRTFs are governed by the Helmholtz equation subject to boundary conditions determined by the listener's individual anatomy~\cite{katz2001boundary}. HRTF upsampling can thus be cast as learning an interpolation operator that maps sparse HRTF measurements at a small number of directions to the complete HRTF field at arbitrary query directions. This operator for the upsampling problem can thus be viewed as function-to-function mapping that must generalize across listeners, each representing a distinct instance of the same underlying acoustic boundary value problem.

We therefore introduce SPARO, a \textbf{spar}se-to-dense neural \textbf{o}perator for HRTF spatial upsampling. SPARO learns a direct mapping from sparse HRTF observations at measured directions to the HRTF field at arbitrary query directions through a learned spherical mode representation that decouples observation and prediction locations. SPARO couples the input and output mappings through a shared learned geometry on the sphere \(S^2\): sparse observations are projected onto spherical latent modes, the resulting mode representations are transformed globally, and the learned mode geometry is used to interpolate the transformed representation at arbitrary query directions. We predict HRTF log-magnitude and ITD on SONICOM from sparse measurements, and evaluate under the LAP24 protocol~\cite{hogg2025listener}. SPARO achieves the lowest error on LSD and ILD at all sparsity levels and is competitive on ITD, pointing to operator learning as a promising direction for personalized spatial audio.

\vspace{-2mm}
\section{Methods}
\label{sec:methods}

% \vspace{-2mm}
\subsection{Preliminary and Notations: HRTF Upsampling
% Head Related Transfer Functions
}

For subject $s$, ear $c \in \{L, R\}$, and source direction 
$\mathbf{d} = (\theta, \phi)$ with azimuth $\theta \in [0, 2\pi)$, and elevation $\phi \in [-\pi/2, \pi/2]$, the HRTF is:
\begin{equation}
    H_s^c(f, \mathbf{d}) = \frac{P_{\text{ear}}^c(f, \mathbf{d})}{P_{\mathrm{free}}(f)},
    \label{eq:hrtf_def}
\end{equation}
where $P_{\mathrm{ear}}^c$ is the sound pressure at the blocked ear-canal
entrance, $P_{\mathrm{free}}$ is the free-field pressure at the head
center \cite{katz2001boundary,gumerov2010computation}, and $f$ denotes frequency. The HRTF is thus a complex-valued transfer function 
encoding both magnitude and phase information as a function of frequency and 
source direction. Its time-domain counterpart is the head-related impulse response 
(HRIR).

Let $\mathbf H_s^c(\mathbf d)\in\mathbb C^F$ denote the HRTF at $F$
frequency bins, and let $\mathcal D$ and $\mathcal S$
denote a dense spatial grid and the training subjects, respectively.
For a test subject $\bar{s}\notin\mathcal S$, HRTFs are available only on a
sparse set $\mathcal D'$; the goal is to estimate
$\mathbf H_{\bar{s}}^c(\mathbf d)$ for all $\mathbf d\in\mathcal D$.

Rather than predict the full complex spectrum, HRTF upsampling methods \cite{masuyama2024niirf,masuyama2025retrieval,zhao2025head} often model log-magnitude
and ITD separately. HRTF magnitude carries direction-dependent spectral
cues important for elevation and front--back localization, while ITD and ILD
provide dominant binaural cues for horizontal localization
\cite{carlini2024auditory, raykar2005extracting}. Minimum-phase HRIRs are reconstructed from the predicted magnitudes and the predicted ITD applied as a fractional delay \cite{kistler1992model}.

% For the purposes of upsampling, it is well-established that full complex-valued HRTF modeling is unnecessary: horizontal localization is primarily governed by the interaural time difference (ITD) and interaural level difference (ILD), while elevation localization depends on spectral coloration in the magnitude response \cite{geronazzo2025technical, carlini2024auditory}. It therefore suffices to model HRTF log-magnitude and ITD separately, reconstruct minimum-phase HRIRs, and apply ITD as a fractional delay \cite{kistler1992model}. This decomposition, adopted by recent methods \cite{masuyama2024niirf, masuyama2025retrieval, zhao2025head}, captures perceptually relevant localization cues while avoiding the complexity of full complex spectrum prediction. Accordingly, we model the HRTF log-magnitude and ITD separately and combine their predictions to reconstruct the full HRTF.

% \subsection{Operator Learning}

\vspace{-2mm}
\subsection{Formulating HRTF Upsampling as Operator Learning}

Let $\mathcal D'=\{\mathbf d_i'\}_{i=1}^{M}$ denote the set of $M$
sparse measurement directions. SPARO learns an operator of the form
\begin{equation}
    \mathcal F:
    \left\{
    \bigl(\mathbf d_i',\mathbf u_s(\mathbf d_i')\bigr)
    \right\}_{i=1}^{M}
    \longmapsto
    \mathbf u_s(\mathbf d),
    \qquad \mathbf d\in\mathcal D,
    \label{eq:operator}
\end{equation}
mapping sparse observations of a directional field on $S^2$ to its value
at arbitrary query directions. We apply this formulation both to
the log-magnitude
$\mathbf u_s(\mathbf d)=\log_{10}|\mathbf H_s^c(\mathbf d)|\in\mathbb R^F$
and to the scalar ITD field
$\mathbf u_s(\mathbf d)=\tau_s(\mathbf d)\in\mathbb R$.

\subsection{SPARO Architecture}

To support arbitrary sparse measurement and query locations, SPARO maps the HRTF at observed positions to a fixed set of latent modes, processes the latent representation,
and decodes the result at arbitrary query directions. The overall architecture is shown in Fig.~\ref{fig:placeholder}. Since HRTFs are functions of direction on \(S^2\), we parameterize these latent modes by learned spherical centers, and the prediction at a query direction attends to the processed modes with weights determined by angular proximity to these centers.

\noindent \textbf{Input Embeddings.} Following the branch–trunk construction used in neural operators such as DeepONet \cite{lu2021learning} and LNO~\cite{wang2024latent}, a branch multilayer perceptron (MLP) embeds each observed pair
$(\mathbf d_i',\mathbf u_s(\mathbf d_i'))$, for $i \in \{1,2,\dots, M\}$ as $\mathbf{b}_i \in \mathbb{R}^D$, while a trunk MLP embeds a query direction $\mathbf d_j$ alone as $\mathbf p_j \in \mathbb{R}^D$. Since the branch requires an observed field value but the trunk requires only a direction, the branch is evaluated only at the measured locations, whereas the trunk can be evaluated independently at query directions.

\begin{table*}[!t]
    \centering
    \caption{HRTF upsampling performance on the SONICOM dataset under the
    LAP24 protocol. Results are reported for different numbers of sparse
    HRTF measurements. Lower is better for all metrics.
    Best results are shown in bold.}
    \label{tab:sonicom_results}
    % \setlength{\tabcolsep}{3.3pt}
    % \renewcommand{\arraystretch}{1.08}
     % \scriptsize
    % \resizebox{\textwidth}{!}
    {%
    \begin{tabular*}{\textwidth}{
        @{\extracolsep{\fill}}
        l|cccc|cccc|cccc
        @{}
    }
        \hline
        & \multicolumn{4}{c|}{\textbf{LSD (dB)} $\downarrow$}
        & \multicolumn{4}{c|}{\textbf{ILD error (dB)} $\downarrow$}
        & \multicolumn{4}{c}{\textbf{ITD error ($\mu$s)} $\downarrow$} \\

        \textbf{Method}
        & \textbf{3} & \textbf{5} & \textbf{19} & \textbf{100}
        & \textbf{3} & \textbf{5} & \textbf{19} & \textbf{100}
        & \textbf{3} & \textbf{5} & \textbf{19} & \textbf{100} \\
        \hline

        Nearest Neighbor
        & 8.7 & 8.3 & 5.4 & 3.4
        & 7.6 & 4.8 & 3.0 & 1.4
        & 274.2 & 154.2 & 108.3 & 44.0 \\

        HRTF Selection (ITD) \cite{masuyama2025retrieval}
        & 5.8 & 5.6 & 5.5 & 5.4
        & 1.4 & 1.5 & 1.6 & 1.4
        & 26.3 & 24.5 & 23.2 & 20.0 \\

        NF + LoRA \cite{masuyama2024niirf}
        & 4.7 & 4.6 & 4.1 & 3.8
        & 1.3 & 1.4 & 1.1 & 1.1
        & 28.6 & 24.7 & 14.7 & \textbf{9.1} \\

        RANF \cite{masuyama2025retrieval}
        & 4.4 & 4.6 & 3.6 & 3.0
        & 1.2 & 1.2 & 1.0 & 0.8
        & \textbf{20.5} & \textbf{18.7} & 14.2 & 10.0 \\

        IOA3D \cite{zhao2025head}
        & 5.0 & 5.3 & 3.6 & 2.8
        & 1.3 & 1.4 & \textbf{0.8} & 0.7
        & 31.2 & 30.0 & 30.4 & 25.9 \\

        \hline
        \textbf{Ours}
        & \textbf{4.1} & \textbf{4.1} & \textbf{3.1} & \textbf{2.3}
        & \textbf{1.1} & \textbf{1.0} & \textbf{0.6} & \textbf{0.4}
        & 20.8 & {19.0} & \textbf{14.0} & 10.4 \\

        {{Ours} (single model, all $M$)}
        & 4.1 & 4.1 & 3.2 & 2.8
        & 1.1 & 1.1 & 0.7 & 0.6
        & 21.6 & 19.1 & 15.1 & 12.8 \\
        Ours w/o TTO
        & 4.1 & 4.1 & 3.3 & 2.5
        & 1.1 & 1.0 & 0.8 & 0.4
        &21.0  & 19.1 & 15.2 & 12.7 \\
        \hline

        % \quad w/o test-time optimization
        % & 4.1 & 4.1 & 3.3 & 2.5
        % & 1.1 & 1.0 & 0.8 & 0.4
        % &21.1  & 19.1 & 15.2 & 12.7 \\

        % \quad w/o local cross-attention
        % & 4.2 & 4.2 & 3.4 & 2.9
        % & 1.1 & 1.1 & 0.7 & 0.5
        % & 21.0 & 20.4 & 15.2 & 12.1 \\

        % \hline
    \end{tabular*}%
    }
\end{table*}

\noindent \textbf{Spherical Mode Projection Encoder.} The encoder maps the variable-size set of $M$ observations to a fixed
set of $M_z$ latent modes. Each mode $k$, for $ k \in \{1, 2, \dots , M_z\}$, is associated with a
learnable center $\mathbf c_k\in S^2$ and positive temperature $\tau_k$.
Observations are weighted according to their angular similarity
to $\mathbf c_k$:
\begin{equation}
A_{ik}
=
\operatorname{softmax}_{i}
\left(
\frac{\mathbf d_i'^{\top}\mathbf c_k}{\tau_k}
\right),
\quad
\mathbf z_k^0
=
\sum_{i=1}^{M}
A_{ik}\,W_v\mathbf b_i ,
\label{eq:spherical_encoder}
\end{equation}
where $\mathbf d_i'$ and $\mathbf c_k$ are unit Cartesian vectors,
$\mathbf b_i\in\mathbb R^D$ is the branch embedding of
$(\mathbf d_i',\mathbf u_s(\mathbf d_i'))$, and $W_v$ is a learned value
projection. The softmax is taken over the $M$ available observations
independently for each mode, so each $\mathbf z_k^0$ is a weighted aggregate
of the observed field values associated with a learned region of the sphere.
The centers are initialized uniformly on a Fibonacci lattice on $S^2$ and learned
jointly with the network, while $\tau_k$ controls the angular concentration
of each mode.

\noindent \textbf{Transformer Blocks.} The latent representation
$\mathbf Z^0\in\mathbb R^{M_z\times D}$ is then processed by $L$ standard
Transformer blocks \cite{vaswani2017attention}, each comprising multi-head
scaled dot-product self-attention and a position-wise MLP with residual
connections and layer normalization. Since this
processing is performed only over the $M_z$ latent modes, its cost is
independent of the number of output query directions.

% \textbf{Transformer Blocks.}
% The latent representation $\mathbf{Z}^0 \in \mathbb{R}^{M_z \times D}$ 
% is processed by $L$ Transformer blocks, each consisting of multi-head 
% scaled dot-product self-attention followed by a 
% position-wise MLP, both with residual connections and layer normalization:
% %
% \begin{align}
%     \mathbf{Z}^l &= \mathrm{MLP}\!\left(\mathrm{LN}\!\left(
%     \hat{\mathbf{Z}}^l\right)\right) + \hat{\mathbf{Z}}^l, \\
%     \hat{\mathbf{Z}}^l &= \mathrm{Attention}\!\left(\mathrm{LN}\!\left(
%     \mathbf{Z}^{l-1}\right)\right) + \mathbf{Z}^{l-1}.
%     \label{eq:transformer}
% \end{align}
% %
% All operations in Eq.~\eqref{eq:transformer} are performed in the latent 
% space of dimension $M_z \times D$, with complexity $\mathcal{O}(M_z^2 D)$ 
% independent of the number of output query directions $N$.

\noindent \textbf{Spherical Mode Projection Decoder.}
To obtain the field value at an arbitrary query direction
$\mathbf d_j\in S^2$, the decoder weights the transformed latent modes
according to their angular similarity to the same learned spherical centers:
\begin{equation}
D_{jk}
=
\operatorname{softmax}_{k}
\left(
\frac{\mathbf d_j^\top \mathbf c_k}{\tau_k}
\right),
\qquad
\mathbf r_j^{\mathrm{mode}}
=
\sum_{k=1}^{M_z}
D_{jk}\,\widetilde{\mathbf z}_k^L ,
\label{eq:spherical_decoder}
\end{equation}
where $\widetilde{\mathbf z}_k^L$ denotes a learned transformation of the
final latent mode and $\mathbf d_j$ and $\mathbf c_k$ are unit Cartesian
vectors. The softmax is taken over the $M_z$ latent modes independently for
each query, so each query representation is a location-dependent mixture of
the learned modes. 
%A final MLP maps $\mathbf r_j^{\mathrm{mode}}$ to the predicted field value $\widehat{\mathbf u}_s(\mathbf d_j)$.

The encoder and decoder share the same learned spherical mode centers $\mathbf c_k$ and temperatures $\tau_k$. The encoder projects sparse observations onto learned spatial modes on \(S^2\), the Transformer models interactions between these modes, and the decoder interpolates the transformed modes at arbitrary query directions.

\noindent \textbf{Local Cross-attention.}
In the spherical-mode pathway, measured HRTFs influence a query only after
being aggregated into latent modes. To provide a direct
query-to-observation pathway, we apply cross-attention between
the query and the original sparse measurements. Using the trunk embedding
$\mathbf p_j$ of query direction $\mathbf d_j$ as the query, and the branch
embeddings $\{\mathbf b_i\}_{i=1}^{M}$ as keys and values, \begin{align*}
&\mathbf r_j
=
\mathbf r_j^{\mathrm{mode}}
+
\operatorname{LN}\!\left(
\operatorname{CrossAttn}
(\mathbf p_j,\{\mathbf b_i\}_{i=1}^{M})
\right),\\
&\widehat{\mathbf u}_s(\mathbf d_j)
=
\operatorname{OutMLP}(\mathbf r_j).
\end{align*}
Here, $\widehat{\mathbf u}_s(\mathbf d_j)$ is the final prediction at query
direction $\mathbf d_j$. Each prediction thus attends directly to
available measurements, in addition to the representation through the
learned spherical modes.

\subsection{Implementation and Test-time Personalization}

\textbf{HRTF Reconstruction}.
For HRTF upsampling, we train two instances of the SPARO model: one for HRTF log-magnitude and one for ITD. The magnitude model maps sparse observations \(\{(\mathbf d_i', \log |\mathbf{H}^c_s(\mathbf d_i')|)\}_{i=1}^M\) to the log-magnitude response at arbitrary query directions. A second SPARO model predicts the scalar ITD field \(\tau_s(\mathbf d)\), taking sparse ITD observations \(\{(\mathbf d_i',\tau_s(\mathbf d_i'))\}_{i=1}^M\) as input. The two models are trained independently and their predictions are combined at inference to reconstruct the full HRTF.

% A separate model with identical architecture is trained to predict 
% $\tau_s(\mathbf{d}) \in \mathbb{R}$ at each query direction, taking as 
% input sparse ITD observations $\{(\mathbf{d}_i', \tau_s(\mathbf{d}_i'))
% \}_{i=1}^{M}$. The two models are trained independently and combined at 
% inference to produce the full HRTF reconstruction.

\noindent \textbf{Right Ear Handling}.
Following \cite{zhang2023hrtf}, we represent right-ear measurements in a mirrored coordinate system. Specifically, right-ear azimuths are reflected as \(\theta \mapsto (360^\circ-\theta)\bmod 360^\circ\), allowing both ears to share a common left-ear convention and effectively doubling the number of training examples. This assumes approximate bilateral symmetry after reflection about the median plane.

% To exploit left-right HRTF symmetry, right ear samples are presented to 
% the model in a mirrored coordinate system: azimuths are reflected as 
% $\theta \mapsto (360° - \theta) \bmod 360°$, so that the model always 
% receives input in a canonical left-ear convention. The mirroring is applied 
% only to model-space coordinates; observation indices are computed from the 
% physical grid and are therefore identical for both ears, consistent with 
% both ears being measured at the same $M$ physical directions per subject.

\noindent \textbf{Test-time Adaptation}. For each test subject, we first encode the sparse observations to obtain the initial latent representation \(Z_0\). We then perform a short test-time optimization (TTO) using only the observed measurements, jointly updating \(Z_0\) and the final output MLP head while keeping the latent Transformer blocks and all remaining model parameters fixed. The adaptation objective combines relative \(L^2\) error at the observed directions with penalties that keep the optimized latent and output-layer parameters close to their pretrained values to prevent overfitting on measurements. After adaptation, the optimized latent is decoded at all query directions to obtain the final HRTF prediction. The loss at test-time is thus: 
\begin{equation*}
\mathcal{L}_{\mathrm{TTO}} =
\frac{\left\|\hat{\mathbf{u}}_{\mathcal{D}'}
-\mathbf{u}_{\mathcal{D}'}\right\|_2}
{\left\|\mathbf{u}_{\mathcal{D}'}\right\|_2}
+\lambda\left\|\mathbf{Z}-\mathbf{Z}_0\right\|_2^2
+\lambda\left\|\theta_{\mathrm{out}}
-\theta_{\mathrm{out}}^{(0)}\right\|_2^2 ,
\end{equation*}
where \(\mathbf Z_0\) is the latent representation produced by the pretrained encoder, \(\mathbf Z\) its adapted value; \(\theta_{\mathrm{out}}^{(0)}\) and \(\theta_{\mathrm{out}}\) are the pretrained and adapted output-MLP parameters. We set $\lambda=1$ in all experiments.

\vspace{-3mm}
\section{Experiments and Results}
\label{sec:results}

\subsection{Experimental Setup}

Models are trained for log-magnitude and ITD with $1024$ latent modes, latent dimension $128$, and $4$ Transformer blocks, optimized for 200 epochs with Adam \cite{kingma2014adam} (learning rate 
$10^{-3}$ with 20-epoch warmup and cosine decay, batch size 1) using relative \(L^2\) loss on a single A6000 GPU. We report results both for models trained at a single sparsity level and for a single model trained on all sparsity configurations jointly, evaluated at every level without retraining.
\iffalse
\begin{figure}[h]
    \centering
    \includegraphics[width=\linewidth, trim={0cm 1.5cm 0 0},clip]{lap_grid.png}
    \vspace{-5mm}
    \caption{Sparse measurement positions specified in the Listener Acoustic Personalization Challenge.}
    \label{fig:grids}
\end{figure}
\fi

% \begin{figure}[h]
%     \centering
%     \includegraphics[width=0.32\linewidth, trim={0cm 0cm 0 0},clip]{method_comparison_sp3_subj187.png} \hfill
%     \includegraphics[width=0.32\linewidth, trim={0cm 0cm 0 0},clip]{method_comparison_sp19_subj184.png}\hfill
%      \includegraphics[width=0.32\linewidth, trim={0cm 0cm 0 0},clip]{method_comparison_sp100_subj194.png}
%     \vspace{-2mm}
%     \caption{HRTF magnitude predicted by different methods.}
%     \label{fig:spec}
% \end{figure}

\textbf{Evaluation:} Evaluation uses the three objective metrics defined in the LAP24 challenge~\cite{hogg2025listener}. The LSD measures spectral 
fidelity between the upsampled and reference HRTFs:
\begin{equation}
\mathrm{LSD}
=
\frac{1}{2N}
\sum_{c\in\{L,R\}}
\sum_{n=1}^{N}
\sqrt{
\frac{1}{B}
\sum_{b=1}^{B}
\left(
20\log_{10}
\frac{
|\widehat{H}^{\,c}(f_b,\mathbf d_n)|
}{
|H^{c}(f_b,\mathbf d_n)|
}
\right)^2
},
\label{eq:lsd}
\end{equation}
where $H^{c}(f_b,\mathbf d_n)$ and
$\widehat{H}^{\,c}(f_b,\mathbf d_n)$ are the reference and reconstructed
HRTFs for ear $c$, frequency bin $f_b$ and direction $\mathbf d_n$,
respectively. $B$ is the number of frequency bins and $N$ is the number of
evaluation directions. ILD and ITD errors are computed as in \cite{hogg2025listener}, as mean absolute errors (MAE) in dB and $\mu$s respectively.

\iffalse
The ITD error quantifies timing discrepancies between ears across directions:
%
\begin{equation}
    \text{ITD}_{\text{err}} = \frac{1}{N} \sum_{n=1}^{N} \left| 
    \left( t^L(\mathbf{d}_n) - t^R(\mathbf{d}_n) \right) - 
    \left( \hat{t}^L(\mathbf{d}_n) - \hat{t}^R(\mathbf{d}_n) \right) 
    \right|,
    \label{eq:itd}
\end{equation}
%
where $t^L(\mathbf{d}_n)$ and $t^R(\mathbf{d}_n)$ are the time-of-arrival estimates 
for the left and right ears from the reference HRTF, and $\hat{t}^L$, $\hat{t}^R$ 
are the corresponding upsampled estimates. ITD error is expressed in microseconds ($\mu$s).

The ILD error captures level differences between ears across frequency and direction:
%
\begin{align*}
    &\text{ILD}_{\text{err}} = 
    \\
    &\frac{1}{N} \sum_{n=1}^{N} \frac{1}{B} \sum_{b=1}^{B}
    \left| 
    20 \log_{10} \frac{|H^L(f_b, \mathbf{d}_n)|}{|H^R(f_b, \mathbf{d}_n)|}
    - 20 \log_{10} \frac{|\hat{H}^L(f_b, \mathbf{d}_n)|}{|\hat{H}^R(f_b, \mathbf{d}_n)|}
    \right|,
    \label{eq:ild}
\end{align*}
%
where superscripts $L$ and $R$ denote left and right ear channels respectively. 
ILD error is expressed in decibels (dB).  
\fi

\vspace{-3mm}
\subsection{Experiments on the Measured SONICOM Dataset}
We train and evaluate on SONICOM \cite{engel2023sonicom} following the same setting as in RANF \cite{masuyama2025retrieval}. The dataset provides 793 measured directions per subject at 48 kHz; free-field compensation is applied with a minimum-phase filter. Subject P0079 is excluded because of atypical ITD behavior, leaving 199 subjects. When comparing with \cite{masuyama2025retrieval}, we use the same 20 test subjects; they hold out a validation set for checkpoint and hyperparameter selection, while we train for a fixed number of epochs with a fixed hyperparameter set on the remaining 179 subjects. Upsampling is evaluated at four sparsity levels, $M \in \{3, 5, 19, 100\}$, using the LAP24 measurement configurations.

\begin{figure}[t]
    \centering
    \includegraphics[width=\linewidth, trim={0cm 0cm 0 0},clip]{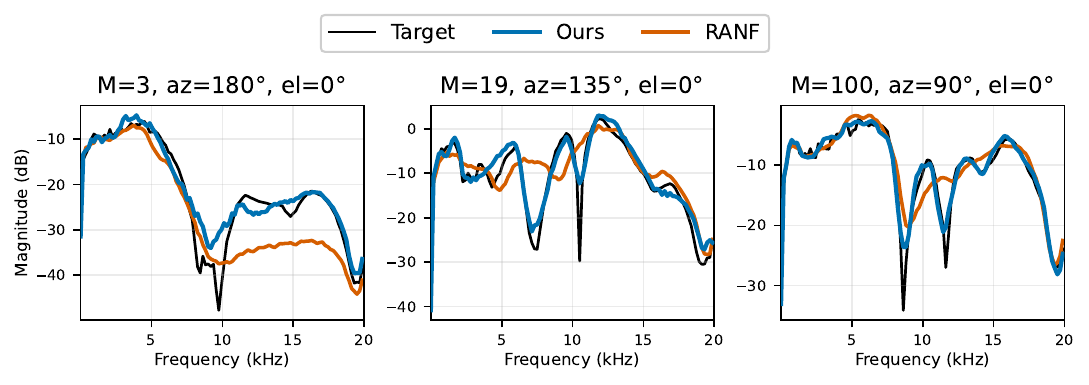} 
    \vspace{-7mm}
    \caption{HRTF magnitude predicted by SPARO and RANF at different sparsity levels and directions. \textcolor{red}}
    \label{fig:spec}
    \vspace{-5mm}
\end{figure}

Tab.~\ref{tab:sonicom_results} reports performance against various baselines, including RANF, which reported the best performance across metrics in \cite{masuyama2025retrieval}.  Following the setup in \cite{masuyama2025retrieval}, metrics are evaluated only at unmeasured directions \(\mathcal{D}\setminus \mathcal{D}'\). Baseline results for nearest neighbor, HRTF selection, NF+LoRA and RANF are taken from \cite{masuyama2025retrieval}. We retrained IOA3D \cite{zhao2025head} under the same protocol as our method.
SPARO achieves the lowest LSD at every sparsity level, and improves on the
best competing method by up to 18\% (2.3\,dB against 2.8\,dB at
$M{=}100$), with consistent gains of 7--14\% at $M{=}3$, $5$ and $19$.
ILD error is also lowest throughout, with a best-case improvement of
43\% at $M{=}100$. ITD error is competitive with the best-performing methods at
all sparsity levels. The single model trained across all 
$M$ matches the per-$M$ models at 
$M=3$ and $5$ and remains competitive at higher sparsity levels.
Fig.~\ref{fig:spec} compares predicted and reference magnitude
spectra at various directions; the reconstruction tracks the target spectral features even at $M{=}3$.

\vspace{-3mm}
\subsection{Experiments on the Simulated HUTUBS Dataset}

For HUTUBS-simulated \cite{brinkmann2019cross} we compare against SUpDEq \cite{arend2023magnitude} as a non-neural spherical-harmonic baseline, following the sampling and evaluation setup in \cite{arend2023magnitude}, where the spherical-harmonic order is prescribed by the sampling grid. The LAP24 SONICOM configurations do not correspond to such grids, so applying SUpDEq there would require additional choices of interpolation order and regularization. The observed directions are Lebedev grids ranging from 6 to 170 directions (degree $Q=1, \dots, 10$) and a degree-29 Fliege target grid containing 900 directions. We use subjects 2--95, excluding the KEMAR heads (subjects 1 and 96), with subjects 86--95 held out for testing and the remaining subjects used for training.

Tab.~\ref{tab:supdeq_lno_all} evaluates SPARO on simulated HUTUBS HRTFs. Since the Lebedev observation grids are not subsets of the Fliege target grid, metrics are reported over all 900 target directions. SPARO achieves lower LSD at every Lebedev degree, with the advantage increasing in the sparser regimes. SPARO reaches 3.39 dB from only six observed directions, a level the baseline does not attain until approximately \(Q=4\) (\(M=38\)), corresponding to roughly $4$--$6\times$ reduction in the number of required measurements. 
% Per-subject standard deviations are comparable to or smaller than those of SUpDEq at all degrees, indicating that the improvement is consistent across test subjects rather than driven by a few. 
SPARO achieves lower LSD than SUpDEq and RANF for all ten held-out subjects at every sampling level. As on SONICOM, the single model trained across all $M$ is strongest in the sparse regime.

\iffalse
Two-sided paired Wilcoxon signed-rank tests across subjects yield \(p=0.00195\) at each level, with all comparisons remaining significant after Holm correction (\(p_{\mathrm{adj}}=0.0195<0.05\)).
\fi

% Fig. shows LSD as a function of azimuth and
% elevation at $M{=}6$ and $M{=}170$, with the measured directions
% overlaid. Error concentrates away
% from the measurement locations at lower elevations.

\begin{table}
\centering
\caption{HRTF upsampling on simulated HUTUBS dataset. LSD (dB, mean\,$\pm$\,std over the 10 test
subjects) over all 900 target directions with Lebedev degree $Q$ and $M$ observations. Results are reported for SPARO trained separately for each $M$ and
across all $M$.}
\label{tab:supdeq_lno_all}
\setlength{\tabcolsep}{4pt}
\resizebox{\columnwidth}{!}{%
\begin{tabular}{rrcccc}
\toprule
$Q$ & $M$ & SUpDEq & RANF & SPARO & SPARO{{ (all $M$})} \\
\midrule
1  & 6   & 5.32\,$\pm$\,0.26 & 4.04\,$\pm$\,0.25 & 3.39\,$\pm$\,0.31 & \textbf{3.32\,$\pm$\,0.30} \\
2  & 14  & 4.38\,$\pm$\,0.26 & 3.49\,$\pm$\,0.34 & 2.91\,$\pm$\,0.25 & \textbf{2.83\,$\pm$\,0.24} \\
3  & 26  & 3.70\,$\pm$\,0.21 & 3.26\,$\pm$\,0.40 & 2.62\,$\pm$\,0.23 & \textbf{2.57\,$\pm$\,0.21} \\
4  & 38  & 3.26\,$\pm$\,0.21 & 3.07\,$\pm$\,0.37 & \textbf{2.38\,$\pm$\,0.18} & 2.37\,$\pm$\,0.19 \\
5  & 50  & 2.89\,$\pm$\,0.21 & 2.97\,$\pm$\,0.40 & \textbf{2.18\,$\pm$\,0.17} & 2.23\,$\pm$\,0.17 \\
6  & 74  & 2.63\,$\pm$\,0.17 & 2.82\,$\pm$\,0.35 & \textbf{2.10\,$\pm$\,0.18} & 2.15\,$\pm$\,0.17 \\
7  & 86  & 2.34\,$\pm$\,0.16 & 2.76\,$\pm$\,0.34 & \textbf{1.93\,$\pm$\,0.15} & 2.06\,$\pm$\,0.16 \\
8  & 110 & 2.10\,$\pm$\,0.13 & 2.77\,$\pm$\,0.30 & \textbf{1.79\,$\pm$\,0.15} & 2.01\,$\pm$\,0.16 \\
9  & 146 & 1.90\,$\pm$\,0.11 & 2.68\,$\pm$\,0.31 & \textbf{1.66\,$\pm$\,0.13} & 1.95\,$\pm$\,0.16 \\
10 & 170 & 1.76\,$\pm$\,0.11 & 2.72\,$\pm$\,0.28 & \textbf{1.56\,$\pm$\,0.12} & 1.93\,$\pm$\,0.16 \\
\bottomrule
\end{tabular}
}
\end{table}
\vspace{-2mm}

\subsection{Ablations}
Tab.~\ref{tab:hutubs_ablation} evaluates local cross-attention (LCA) and test-time optimization (TTO) on simulated HUTUBS. Removing either degrades performance across most sampling levels. TTO improves LSD at every sampling level on HUTUBS and also shows improvements on SONICOM (Tab.~\ref{tab:sonicom_results}). However, SPARO without it already outperforms baselines at all but the densest HUTUBS level. LCA's benefit grows with \(M\), suggesting it is greatest when enough observations are available for direct query-to-measurement interaction.

\begin{table}[t]
    \centering
    \caption{Ablation on simulated HUTUBS. LSD (dB) with local cross-attention
(LCA) and test-time optimization (TTO) removed.  LCA gains increase with M; TTO helps at every level.
%Both components contributeacross sampling levels.
}
    \label{tab:hutubs_ablation}
    \setlength{\tabcolsep}{5pt}
    \renewcommand{\arraystretch}{1.08}
    \begin{tabular}{l ccccc}
        \toprule
        \textbf{Variant}
        & $M{=}6$ & $M{=}26$ & $M{=}50$ & $M{=}110$ & $M{=}170$ \\
        \midrule

        \textbf{SPARO}
        & 3.39 & \textbf{2.62} & \textbf{2.18} & \textbf{1.79} & \textbf{1.56} \\

        \quad w/o LCA
        & \textbf{3.37} & 2.71 & 2.47 & 2.26 & 2.22 \\

        \quad w/o TTO
        & 3.51 & 2.84 & 2.41 & 2.04 & 1.83 \\
        \bottomrule
    \end{tabular}
    \vspace{-3mm}
\end{table}

\vspace{-3mm}
\subsection{{Computational Efficiency }}

Tab.~\ref{tab:efficiency} compares inference cost against RANF, measured on an A6000 GPU over 793 target directions. SPARO's test-time cost is independent of 
$M$ (0.56–0.59 s with TTO, 12 ms without), whereas RANF's grows from 8.7 s at 
$M=3$ to 207.2 s at 
$M=100$, giving speedups of 15–373$\times$. RANF adapts low-rank subject-specific parameters over 500 epochs of backpropagation through the full network and all retrieved subjects. SPARO's TTO updates only the latent representation and output head for 20 steps.
 
\iffalse
\begin{table}[h]
    \centering
    \caption{Computational efficiency comparison with RANF.
    Speedup is computed relative to RANF. End-to-end test time includes subject-specific adaptation and dense HRTF magnitude reconstruction.}
    \label{tab:efficiency}
    \setlength{\tabcolsep}{5pt}
    \renewcommand{\arraystretch}{1.08}
    \begin{tabular}{lccc}
        \hline
        \textbf{Method}
        & \textbf{Training (s)}
        & \textbf{Inference (s)}
        & \textbf{Parameters} \\
        \hline
        RANF
        & 0.317
        & 26.5
        & $\sim$1.23M \\

        \textbf{Ours}
        & \textbf{0.021}
        & \textbf{0.32}
        & {\textbf{$\sim$0.88M}} \\
        \hline
        \textbf{Improvement}
        & $\mathbf{15.2\times}$
        & $\mathbf{82.9\times}$
        & $\mathbf{1.4\times}$ \\
        \hline
    \end{tabular}
\end{table}
\fi

\vspace{-3mm}
\begin{table}[h]
    \centering
    \caption{Per-subject inference cost, including magnitude and ITD for both methods. Speedup is relative to SPARO with TTO.
    Parameters: 1.73M (SPARO, two networks) vs.\
    0.82M shared + 2,048 per listener (1.23M total, RANF).}
    \label{tab:efficiency}
    \setlength{\tabcolsep}{5pt}
    \renewcommand{\arraystretch}{1.08}
    \begin{tabular}{lcccc}
        \hline
        $M$ & SPARO w/o TTO & SPARO & RANF & Speedup \\
        \hline
        3   & 0.012\,s & 0.59\,s & 8.7\,s   & \textbf{15$\times$} \\
        19  & 0.012\,s & 0.56\,s & 29.3\,s  & \textbf{53$\times$} \\
        100 & 0.012\,s & 0.56\,s & 207.2\,s & \textbf{373$\times$} \\
        \hline
    \end{tabular}
\end{table}
\vspace{-3mm}

\section{CONCLUSION}
\label{sec:conclusion}
\vspace{-5pt}
We formulated HRTF spatial upsampling as operator learning
and introduced SPARO, which maps sparse measurements to arbitrary query directions through a learned spherical mode representation.
SPARO achieves the lowest LSD and ILD error among the compared methods at all four SONICOM sparsity levels, with competitive ITD and substantially lower test-time cost than neural-field methods. Results on HUTUBS further support operator learning as an effective approach for personalized HRTF reconstruction.

% We formulate sparse HRTF upsampling as an operator learning problem and introduce SPARO, which maps sparse measurements to arbitrary query directions through a learned spherical mode representation. 
\newpage

\bibliographystyle{IEEEtran}
\bibliography{IEEEabrv,strings,refs}

\end{document}